\documentclass[twocolumn, prb, showpacs, superscriptaddress]{revtex4-1}
\usepackage{amsmath,amssymb,epsfig,color}
\usepackage{graphicx}
\usepackage{dcolumn}
\usepackage{bm}
\usepackage{color}
 \usepackage{braket}

\begin{document}

\title{Three-site transition-metal clusters: going from localized electrons to molecular orbitals}

\author{Evgenia V. Komleva}
\affiliation{M.N. Mikheev Institute of Metal Physics UB RAS, 620137, S. Kovalevskaya str. 18, Ekaterinburg, Russia}
\author{Daniel I. Khomskii}
\affiliation{II. Physikalisches Institut, Universit{\"a}t zu K{\"o}ln, Z{\"u}lpicher Stra{\ss}e 77, D-50937 K{\"o}ln, Germany}
\author{Sergey V. Streltsov}
\affiliation{M.N. Mikheev Institute of Metal Physics UB RAS, 620137, S. Kovalevskaya str. 18, Ekaterinburg, Russia}
\affiliation{Ural Federal University, Mira St. 19, 620002 Ekaterinburg, Russia}

\begin{abstract}
A recently synthesised series of isostructural compounds Ba$_4$NbTM$_3$O$_{12}$ (TM = Mn, Rh and Ir) with transition-metal trimers in a face-sharing geometry makes it possible to examine a tendency to the molecular orbitals (MO) formation going from $3d$ to $5d$ transition metal ions. Our $ab$ $initio$ calculations of electronic and magnetic properties describe the experimental findings and demonstrate gradual transition from the picture of localized electrons for Mn to the MO picture for Rh and especially for Ir. We also show that the often used criterion, according to which the
metal-metal distance in a compound shorter than in a respective metal
always gives MO, may break down in some cases.
\end{abstract}    

\date{\today}

\maketitle

\section{Introduction}

Transition metal (TM) compounds have been intensively studied for the last decades and they continue to attract much interest nowadays \cite{Khomskii-book,UFN,Imada1998}.  In some of them electrons behave as mostly localized at the TM sites, while in others they turned out to be completely delocalized, band-like. Moreover, there can be transitions, when electrons change degree of the localization under an external perturbation (pressure, temperature etc.)\cite{Imada1998}. Recently a new class of such materials, so called cluster Mott insulators, with TM forming by well-defined magnetic clusters, attracted considerable attention. In this case electrons are localized not on a single TM site, but on a group of tightly bound ions. One can often consider such systems as a collection of ``molecules'' with significant electron hopping within such clusters, and they may often be described by molecular orbitals (MO).

In this situation nature of electrons is twofold: they {\it simultaneously} behave as itinerant within a ``molecule'' and localized if one considers hopping between such clusters. These hoppings may still be rather small, so that a solid built of such clusters would still behave as a Mott insulator - but with electrons localised not on TM ions, but on such clusters. In effect, these systems on the one hand are very similar to conventional strongly correlated systems: they can be Mott insulators or experience metal-insulator transitions, be magnetic (but with the moment localized not on TM ions as usual, but on such ``molecules'') or even superconducting~\cite{Abd-Elmeguid-2004}. But besides that, in this situation  we can also have extra effects due to  the appearance of internal degrees of freedom inside these clusters. Such ``molecules'' or clusters can be treated as structural units of a system that determine physical properties of a given compound.  

Very typical for such systems is the situation when inside such clusters the metal-metal distances are rather short - often smaller than in respective elemental metals - and thus corresponding electron hopping, $t$ may be of the order or even larger than the intra-atomic parameters such as the Hubbard's $U$, Hund's coupling $J_H$ and spin-orbit coupling (SOC) constant $\lambda$.
In general, $3d$ electrons are usually more localized, have smaller intersite hopping, but larger Hubbard repulsion than $4d$ and $5d$, thus they may behave as more localized, and one might expect better conditions for the formation of MO states in $4d$ and $5d$ systems than in $3d$ ones (although of course such MO states are not excluded also in some $3d$ systems). Another important factor could be the influence of SOC on the properties of these compounds, which also change regularly when we go from the $3d$ to $4d$ and $5d$ systems.

In this sense it would be very interesting to consider the tendency for the MO formation for the series of similar materials when  we move ``vertically'' in the periodic table. Recently a number of Ba$_4$NbTM$_3$O$_{12}$ compounds, where TM are Mn, Ru, Rh or Ir, with similar structural TM trimers were synthesised by Loi T. Nguyen et al.\cite{Nguyen-2018,Nguyen-2019,Nguyen-2019-Ir}. These materials, having practically the same crystal structure, show very different magnetic properties. There is rather large effective magnetic moment of 4.82~$\mu_B$/f.u.\cite{Nguyen-2019} in Ba$_4$NbMn$_3$O$_{12}$, while these values for Rh- and Ir-based systems are 1.48 $\mu_B$/f.u. and 0.80 $\mu_B$/f.u., respectively\cite{Nguyen-2019-Ir}. Ba$_4$NbMn$_3$O$_{12}$ exhibits long-range magnetic order below 42 K, but if replaced by Rh ordering temperature drops to 1.5 K, while iridate remains nonmagnetic down to the lowest temperatures and was suggested to be a spin liquid. The Curie-Weiss temperatures for Ba$_4$NbTM$_3$O$_{12}$ are -4 K, -23 K and -13 K for Mn, Rh and Ir, respectively, signalling rather small inter-cluster exchange. Furthermore, these systems along with the other hexagonal perovskites are promising compounds for finding new quantum materials \cite{Cava-review}.

These TM ions belong to different periods, $3d$, $4d$ and $5d$. This gives us the possibility to examine the MO-formation tendency going from $3d$ to $5d$ transition metal ions in the same class of materials.  This is the main goal of the present investigation which we carry out using the \textit{ab initio} calculations.

In this paper we analyze electronic structure and magnetic properties of Ba$_4$NbIr$_3$O$_{12}$ and Ba$_4$NbRh$_3$O$_{12}$. Additionally, to complete the picture, we discuss and compare these results with  our previous findings for  Ba$_4$NbMn$_3$O$_{12}$, presented as a short letter in Ref. \cite{Streltsov-2018}. The results obtained indeed confirm our general expectations and demonstrate gradual transition from the picture of localized electrons for Mn to the MO picture for Rh and especially for Ir. Our results also show that the ``rule of thumb'' according to which MO states are realized when the metal-metal distance in a compound is of the order or smaller than in respective metal, may break down, especially for some 3d systems.


\section{Calculations}

We performed \textit{ab-initio} calculations of electronic structure of Ba$_4$NbIr$_3$O$_{12}$ and Ba$_4$NbRh$_3$O$_{12}$ using mostly Wien2k \cite{wien2k}. Crystal orbital Hamiltonian occupation (COHP) analysis \cite{COHP} was carried out by LOBSTER\cite{LOBSTER}. Charge densities for particular electronic bands and data for further COHP analysis for all three systems were obtained in Vienna ab initio Simulation Package (VASP) \cite{Vasp}. The exchange-correlation functional was chosen to be in the form proposed by Perdew et $al.$ \cite{pot}. For Wien2k calculations we took a mesh consisting of 300 $k$ points while for VASP simulations 8$\times$8$\times$8 grid was used. VASP plane-wave cutoff energy was 700 eV, for Wien2k calculation the default parameter of $RK_{max}$=7.0 was taken.
\begin{figure}[t]
\includegraphics[width=0.4\textwidth]{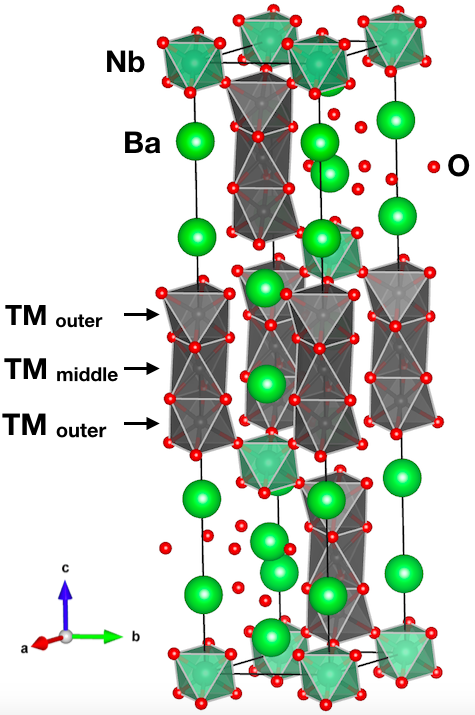}
\caption{\label{label} Crystal structure of Ba$_4$NbTM$_3$O$_{12}$. TM ions (Mn, Rh, Ir) shown in grey are in the oxygen (small red balls) octahedra. Three nearest TMO$_6$ octahedra sharing faces form trimers. The TM ions, which are in the middle of the trimer, are labeled as ``middle'', while ``outer'' are outer TM ions in the trimer. Large green balls are Ba atoms, small green ones in oxygen octahedra are Nb ions.}
\label{Crystal_str}
\end{figure} 

The calculations were performed for different values of Hubbard $U$ and Hund's $J_H$ parameters. For Ir ions $U$ was taken in range 1.5-2.0 eV, whereas $J_H$ belongs to 0.3-0.5 eV \cite{Ir-U} region. For Rh ions $U$ and $J_H$ values vary from 2.5 to 3 eV and from 0.5 to 0.7 eV \cite{Rh-U}, respectively. Results for Mn-contained trimers were reported with $U$ equal to 4.5 eV and $J_H$=0.9 eV \cite{Streltsov-2018, Streltsov-2014}. This illustrates tendency of interaction strength to increase going from the $5d$ to $3d$ elements.

We considered both non-magnetic and magnetically ordered states of these systems with or without including SOC. In magnetic cases atoms were arranged either ferromagnetically (FM) or in a simple antiferromagnetic (AFM) manner ($\uparrow-\downarrow-\uparrow$) in a trimer. The unit cell contains only one trimer.

Crystal structure was taken from Ref.\cite{Nguyen-2018,Nguyen-2019,Nguyen-2019-Ir} and is shown in Fig. \ref{Crystal_str}. We are interested mostly in a building block of a linear trimer of three face-sharing TMO$_6$ octahedra. TM-TM bond lengths in these trimerzied systems and in metals are summarized in Tab.~\ref{Bonds}. There are two crystallographically different positions for TM ions: in the middle of a trimer and the ``outer'' positions. The average valence of TM in such materials is TM$^{3\frac{2}{3}+}$, i.e. it nominally contains two TM$^{4+}$ and one TM$^{3+}$. In principle they could form charge-ordered (CO) state  (most naturally with the TM$^{3+}$ in the middle and TM$^{4+}$ at the outer sites), but structural data \cite{Nguyen-2018,Nguyen-2019,Nguyen-2019-Ir} do not show any indication for that - at least not strong CO.

As Mn and both Ir and Rh belong to different groups of periodic table their electronic configurations are not the same. Ir and Rh trimers nominally contain $d^5$ and $d^6$ ions while Mn cluster is made of $d^3$ and $d^4$ ions. Here it should be stressed that the possibility of direct comparison between the above mentioned compounds is not absolutely evident: being crystallographical equivalent, Mn system has another number of electrons per trimer (10 to be precise) in comparison with Ir- and Rh-based systems with 16 $d$-electrons.

\begin{table}[t]
 \begin{tabular}{c|c|c|c}  
TM  & metal & trimer & difference \\
\hline
\hline 
Mn & 2.734 & 2.469 & 0.265 \\  
Rh & 2.630 & 2.545 & 0.085 \\
Ir & 2.714 & 2.547 & 0.167 \\
 \end{tabular}
	\caption{Metal-metal bond lengths in \AA~in pure metals and in a trimer in Ba$_4$NbTM$_3$O$_{12}$ for TM = Ir, Rh, and Mn.}
		\label{Bonds}
\end{table}
 \begin{figure}[b]
\includegraphics[width=0.48\textwidth]{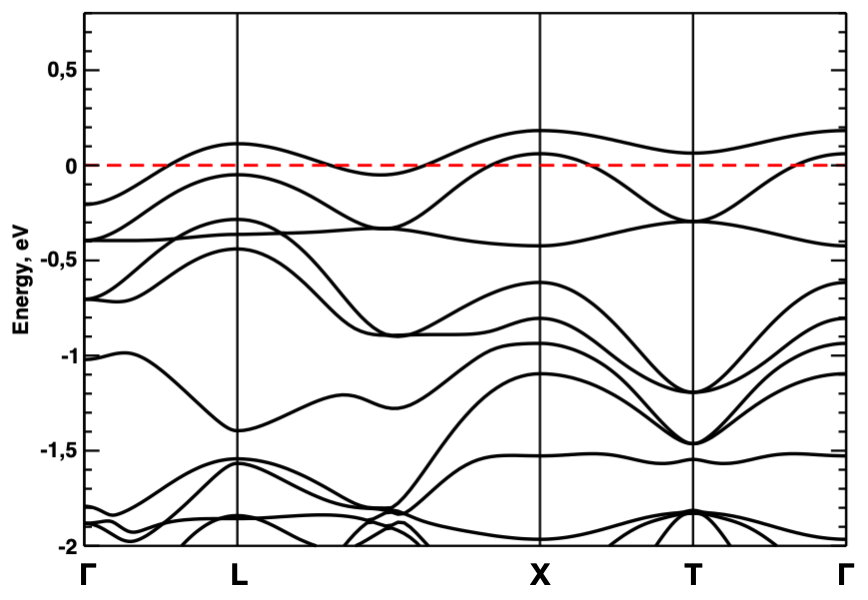}
\caption{\label{label} Band structure of Ba$_4$NbIr$_3$O$_{12}$ as obtained in the non-magnetic GGA calculations. The Fermi energy is at zero. Here and in the others band structure figures high-symmetry points have the following coordinates: L=(0, 0.5, 0), X=(0.5, 0, 0.5) and T=(0.5, 0.5, 0.5).}
\label{Ir_GGA_NM}
\end{figure}


\section{Nonmagnetic GGA results~\label{NM-GGA}}

According to the simple ``rule of thumb'' criterion\cite{UFN} for the MOs formation all these systems can be promising candidates for this phenomenon as typical TM-TM distances in corresponding metals are significantly larger than in the present compounds (see Tab.~\ref{Bonds}). In order to check this hypothesis we start with analysis of nonmagnetic (NM) GGA results.

As an example, in Fig.~\ref{Ir_GGA_NM} the NM GGA band structure for one of the materials: Ba$_4$NbIr$_3$O$_{12}$ is shown. Nine $t_{2g}$ bands of Ir atoms (3 for each) is clearly seen at and below the Fermi level. One may note a significant splitting of these $t_{2g}$ orbitals, which is due to formation of bonding, antibonding, and non-bonding combinations. There is an intrinsic trigonal symmetry in the common face geometry realized in the crystal structure of Ba$_4$NbTM$_3$O$_{12}$ materials. The $t_{2g}$ subshell of a TM ion in a trigonal field is split onto one $a_{1g}$ and two $e_g^\pi$ orbitals.  As a result, for a trimer we have bonding, non-bonding and antibonding $a_{1g}$  states, and similarly for six $e_g^\pi$ orbitals. The $a_{1g}$ orbitals, shown in Fig.~\ref{levels}, are directed to each other and have the strongest direct $d-d$ overlap in a trimer and thus give the largest bonding-antibonding splitting~\cite{Streltsov2012a}. Corresponding one-electron level diagram is presented in the left part of Fig.~\ref{model-sketch}(a).

\begin{figure}[t]
\includegraphics[width=0.45\textwidth]{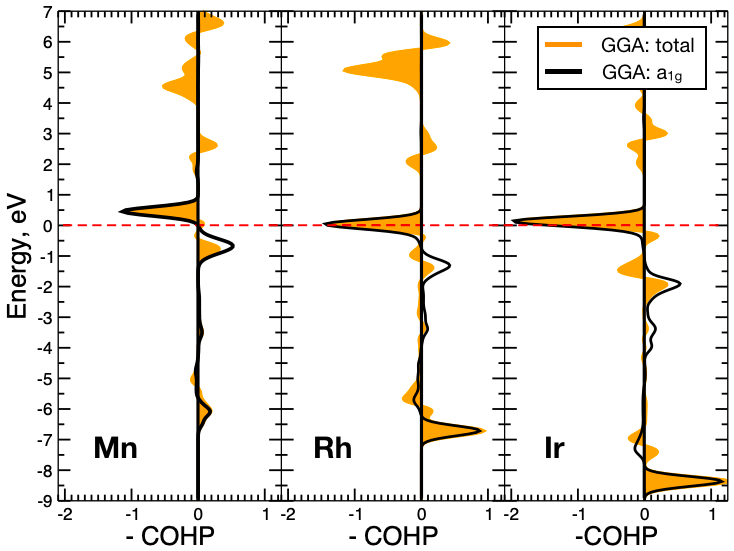}
\caption{\label{label} Results of the COHP analysis for all considered systems in the NM GGA calculations (no $U$, no SOC). Total COHP for the pair of middle and outer TM ions is shown in orange, the COHP plot for only $a_{1g}$ orbitals is colored in black. The Fermi energy is at zero. The fact that total COHP is slightly smaller than COHP for only $a_{1g}$ orbitals is a result of projection and is unimportant for further consideration.
}
\label{COHP}
\end{figure} 

In order to extract model parameters for each system under consideration, we projected NM GGA Hamiltonian onto basis of site-centered $d$ orbitals of a trimer using Wannier90 package\cite{wannier}. The obtained real-space 15$\times$15 Hamiltonian matrix contains information on five $d$ orbitals for each of three TM atoms in a trimer. The Hamiltonian was rotated to the local coordinate system so that each of these main-diagonal blocks corresponding to atomic sites were diagonalized. Off-diagonal elements corresponding to electron hoppings between two orbitals on neighbouring sites along with crystal-field splitting were extracted and presented in Tab.~\ref{parameters}. The largest hoping parameters, $t_{\sigma}$, are between the $a_{1g}$ orbitals looking towards each other in the face sharing geometry, while $t_{\pi}$ are the hoppings between the $e_g^{\pi}$ orbitals.

First, we see that the strongest $a_{1g}$ hopping is for Ir ions and the bonding-antibonding splitting, which is $2\sqrt 2 t_{\sigma}$,~\cite{Streltsov2012a} is larger than intra-atomic exchange $J_H$ and spin-orbit coupling $\lambda$ and of order of Hubbard $U$. Thus one might expect that the MOs regime realizes in Ba$_3$NbIr$_3$O$_{12}$. Situation with Mn is also rather obvious: smaller bonding-antibonding splitting can not compete with much stronger correlation effects in the $3d$ bands, while in the Rh system we are the intermediate regime. Second, we see that going from the $3d$ to $5d$ orbitals the hopping parameter for the $a_{1g}$ orbital changes on $\sim 66$\%. 
\begin{figure}[t]
\includegraphics[width=0.48\textwidth]{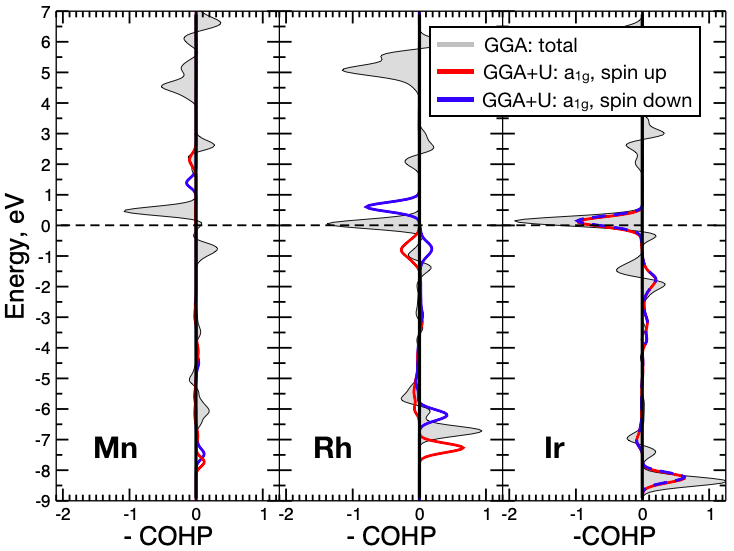}
\caption{\label{COHP+U} Calculated COHP in nonmagnetic GGA for TM-ion pairs (shown in grey) and in magnetic GGA+U for only $a_{1g}$ orbitals (red for positive and blue for negative spin projections). The Fermi energy is at zero.}
\label{COHP+U}
\end{figure} 

We get further insights from the crystal orbital Hamiltonian population (COHP) plot for the pair of outer and middle TM ions calculated in NM GGA and presented in Fig.~\ref{COHP}. The bonding and antibonding states basically manifest themselves as having different signs in the COHP plots\cite{Wills2010}. One may see that results reported in Tab.~\ref{parameters} are in line with COHP analysis. There are two large peaks of different signs in vicinity of the Fermi level, which correspond to the bonding and antibonding $a_{1g}$ bands  (nonbonding band has zero COHP). The splitting between these bands increases going from Mn to Ir as $t_{\sigma}$ does. 

In addition to these two peaks there are large bonding contributions at -6...-8.5 eV. Due to strong covalency effects electrons are localized not on pure atomic $d$ orbitals, but on Wannier functions having rather large contributions from both TM $d$ and ligand $p$ states and this oxygen's contribution increases as we go from $3d$ to $5d$ metals. This is the bonding combination ligand's $2p$ orbitals of $a_{1g}$ symmetry, which is responsible for the large peaks in the COHP at -6...-8.5 eV. The peak intensity strongly increases in Ir comparing to Mn.
 \begin{table*}[t]
    \centering
    \begin{tabular}{c|c|c|c|c|c|c}
        \hline
		\hline
		TM & $t_\sigma$ & $t_\pi$ & $\Delta_{CFS}^{a_{1g}-e_g^\pi}(m)$ & $\Delta_{CFS}^{a_{1g}-e_g^\pi}(o)$ & $\Delta_{CFS}^{t_{2g}-e_g^{\sigma}}(m)$ & $\Delta_{CFS}^{t_{2g}-e_g^{\sigma}}(o)$ \\
		\hline
		Mn  & -0.39 & 0.16 & 0.03 & 0.02  & 2.54 & 2.56  \\
		Rh  & -0.45 & 0.35 & -0.12 & -0.05 & 2.92 & 3.00  \\
		Ir  & -0.65 & 0.44 & -0.34 & 0.07  & 3.82 & 3.22  \\
		\hline
		\hline
    \end{tabular}
\caption{Calculated hopping parameters between the $a_{1g}$ orbitals are $t_\sigma$, $e_g^\pi$ orbitals - $t_\pi$, trigonal crystal-field splitting (CFS) between $a_{1g}$ and $e_g^\pi$ orbitals for middle (m) and outer (o) TM atoms and the crystal-field splitting between centers of $t_{2g}$  and $e_g^\sigma$ subshell in eV. Negative $\Delta_{CFS}^{a_{1g}-e_g^\pi}(m)$ means that the $e_g^\pi$ levels are lower than $a_{1g}$.}
    \label{parameters}
\end{table*}
 \begin{figure}[b]
\includegraphics[width=0.48\textwidth]{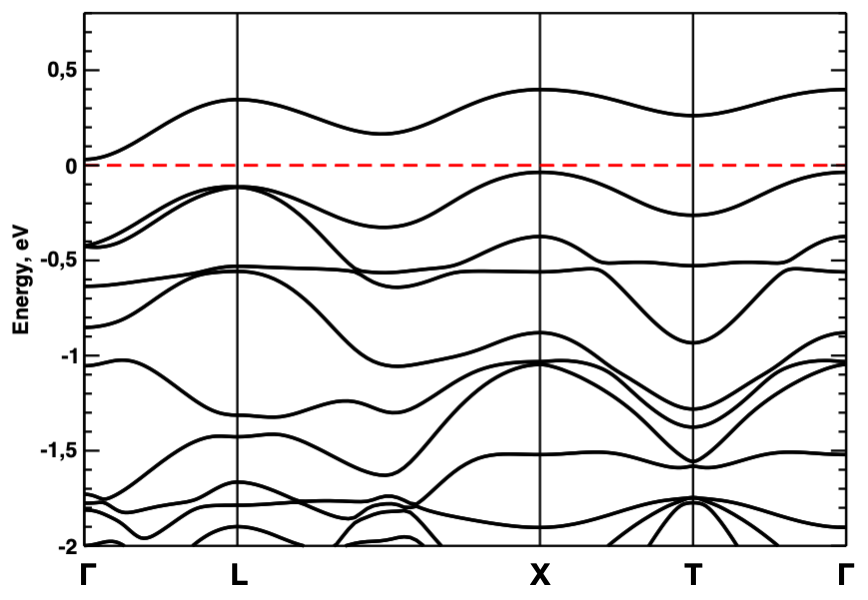}
\caption{\label{Ir_SOC_bands} Calculated band structure for Ba$_4$NbIr$_3$O$_{12}$ in GGA+SOC. Including Hubbard $U$ does not significantly change band structure. The Fermi energy is at zero.}
\end{figure} 

Pairs of bonding and antibonding peaks of the $e_g^{\pi}$ symmetry appears also in vicinity of the Fermi level (the splitting between them is much smaller than for $a_{1g}$) and lower in energy in the region of oxygen $2p$ band (from -5 to -8 eV). The intensities of all peaks as well as the splitting increases going from $3d$ to $5d$ orbitals.

To sum up, one can see that without taking into account Hubbard $U$, Hund's $J_H$ and SOC the TM ions in trimers are strongly coupled with each other and we see formation of the MOs in all of them. In the next step we introduce Hubbard $U$ in these systems to check its influence on the formation of MOs.


\section{GGA+U and GGA+U+SOC results}

First let us consider two limits - $3d$ and $5d$ trimers and then move on to the $4d$ case, that, as we have seen in Sec. III, is expected to be somewhere in between of these two. It would be easier to start with Mn-based material. 

\subsection{Ba$_4$NbMn$_3$O$_{12}$}
 We will not reproduce the detailed calculations and the analysis of this compound which has been presented previously in Ref.\cite{Streltsov-2018}. Nevertheless, some comments on these in light of current discussion has to be done. Introducing reasonable Hubbard $U$ ($\sim$4.5 eV and $J_H$=0.9 eV)  immediately suppresses MO formation in Ba$_4$NbMn$_3$O$_{12}$ and leads to the localized electrons (LE) solution. The COHP analysis shows that correlation effects suppress MOs completely, see Fig.~\ref{COHP+U}. In contrast to NM case depicted in grey here, there are no peaks corresponding to bonding and antibonding bands. One can notice only smeared peak at $\sim$ -8 eV connected again with the Mn-O hybridization.

It looks like the qualitative ``rule of thumb'', mentioned in Sec.~\ref{NM-GGA}, for the formation of MO states is not applicable here. Mn ions in Ba$_4$NbMn$_3$O$_{12}$ have the shortest distances in a trimer and the largest difference from the Me-Me bond length (see Tab.~\ref{Bonds}). Both experimental results and our calculations show that the LE picture much better describes properties of this material, since  it exhibits a long-range magnetic order and significant magnetic moments.

As influence of SOC is not so crucial for $3d$ systems we did not further introduce it in the calculations.
In effect, it turns out that for this $3d$ system intra-atomic interactions play the major role. The detailed analysis of Ref.~\cite{Streltsov-2018} demonstrates that in  Ba$_4$NbMn$_3$O$_{12}$ electrons are indeed localized on Mn ions, which have local magnetic moments coupled by exchange interaction. Now let us consider the opposite limit of $5d$ Ir trimer in Ba$_4$NbIr$_3$O$_{12}$.

\subsection{Ba$_4$NbIr$_3$O$_{12}$}
\begin{figure}[t]
\includegraphics[width=0.49\textwidth]{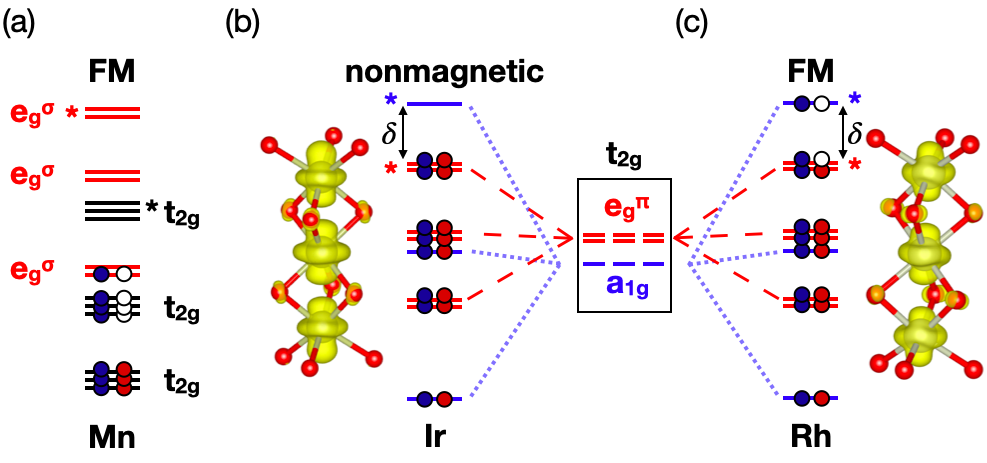}
\caption{\label{label} (a) Level diagram (with both $t_{2g}$ and $e_g^{\sigma}$) for Mn trimer and (b, c) scheme of $t_{2g}$ energy levels (no $e_g^{\sigma}$) for MO model and charge densities of the band above/at the Fermi level for (b) Ir and (c) Rh trimers. Red balls are oxygen atoms. Coloured star denotes antibonding states.}
\label{levels}
\end{figure}

First of all we stress again that direct comparison of Ir and Rh trimers with the above mentioned Mn one is not absolutely evident: being crystallographically equivalent, Mn system has another number of electrons and, correspondingly, another scheme of occupied electron levels. Because of much smaller $t_{2g}-e_g$ crystal-field splitting, see Tab.~\ref{parameters}, in case of Mn also $e_g^\sigma$ orbitals should be taken into account (see Fig.~\ref{levels}). Also correlation effects are strongly enhanced in Mn. In general, for $3d$ systems $J_H$ is typically of 0.8-0.9 eV which is higher than for $4d$ or $5d$, see e.g.~\cite{ChemRev}. But in addition in this case the role of Hund's coupling is increased in comparison with the systems with Ir and Rh: each electron in Mn trimer interacts with 3 others unpaired ones and, consequently, gains energy of not $J_H$ but rather 3$J_H$.

As we have seen, in a simple GGA calculations (no Hubbard $U$, no SOC) Ba$_4$NbIr$_3$O$_{12}$ turned out to be metallic. One may include Hubbard correlations using GGA+U method but it does not help to open the gap at  the Fermi level. However, we can clearly see from the COHP analysis presented in Fig.~\ref{COHP+U} that in contrast to the case of Mn trimer correlation effects do not suppress MO formation.  In the GGA+U calculations the absolute magnitude of peaks, corresponding to bonding and antibonding bands, became slightly smaller than for GGA but still remains significant. For $U$ as large as 1.5 eV and intra-atomic Hund's exchange $J_H$=0.5 eV~\cite{wan2011topological,Dey2016,Streltsov2017,Revelli2019} starting with AFM ordering, the obtained solution is almost nonmagnetic with 0.022 $\mu_B$ on the middle Ir ion and 0.009 $\mu_B$ on the outer ions.

An account of just the spin-orbit coupling in the frameworks of GGA+SOC approximation (even without applying $U$ and $J_H$) immediately leads to the insulating state,  see Fig.~\ref{Ir_SOC_bands}, as in the experiment \cite{Nguyen-2019-Ir}. 
\begin{figure}[t]
\includegraphics[width=0.5\textwidth]{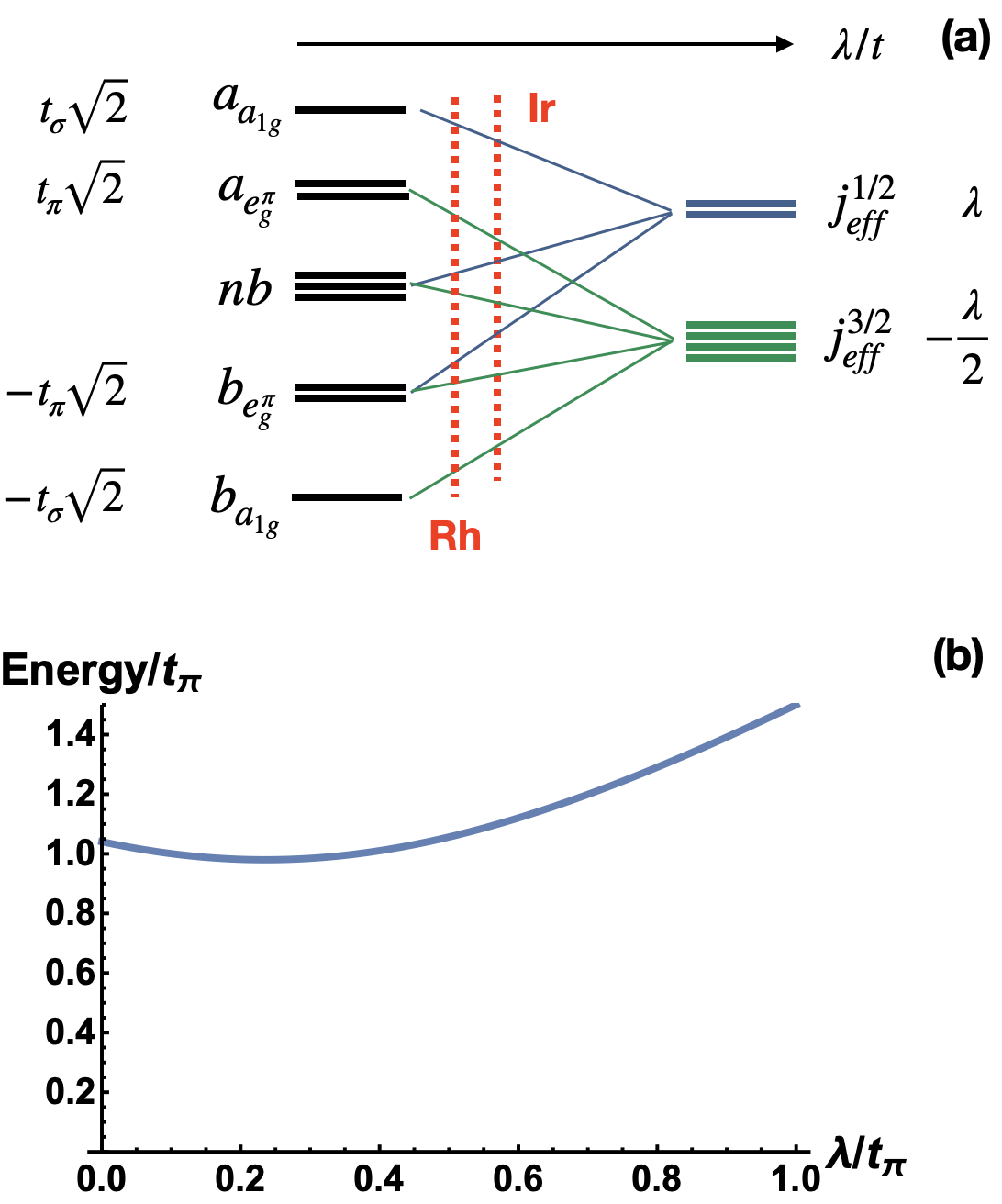}
\caption{\label{model-sketch} (a) Sketch showing how molecular orbitals in a trimer transform to the $j_{eff}^{1/2}$ and $j_{eff}^{3/2}$ levels as strength of the spin-orbit coupling ($\lambda$) changes. Dotted lines illustrate the the situations realized in Ba$_4$NbRh$_3$O$_{12}$  and Ba$_4$NbIr$_3$O$_{12}$, $a_{a_{1g}}$  and $a_{e_g^{\pi}}$ stand for antibonding, $b_{a_{1g}}$  and $b_{e_g^{\pi}}$ for bonding, and $nb$ for non-bonding orbitals. Note that each level in the right part of the diagram is triply (3 ions) degenerate. (b) Results of one-electron calculations of the splitting between with $a_{a_{1g}}$  and $a_{e_g^{\pi}}$ with parameters corresponding to Ba$_4$NbIr$_3$O$_{12}$, see Tab.~\ref{parameters}.}
\end{figure} 

\begin{figure}[t]
\includegraphics[width=0.48\textwidth]{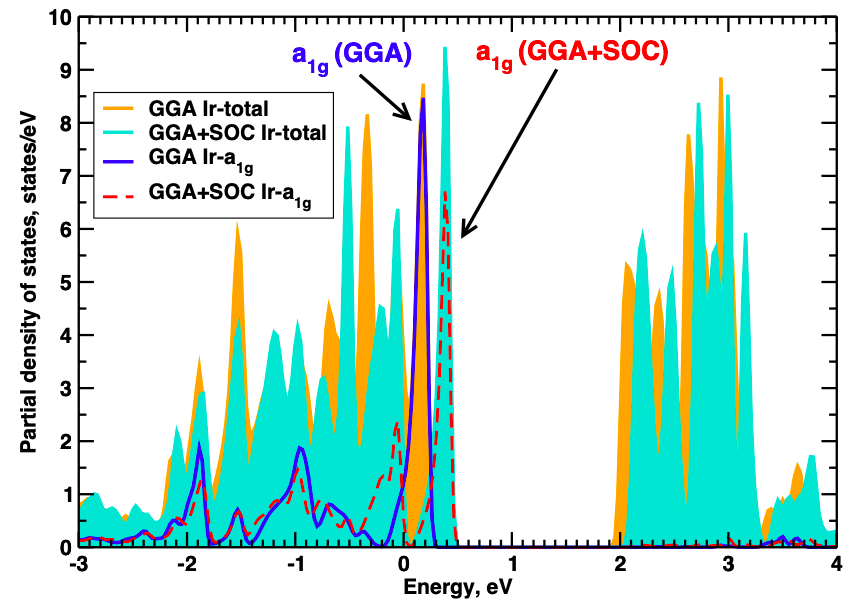}
\caption{\label{Ir_SOC_DOS} Calculated partial density of states for Ir ions in Ba$_4$NbIr$_3$O$_{12}$ in GGA and GGA+SOC. Total and only belonging to $a_{1g}$ orbital Ir DOS are plotted. The Fermi energy is at zero.}
\label{part_dos_Ir}
\end{figure}

The origin of this effect can be understood from the one-electron model calculations for three sites with $t_{2g}$ orbitals only, where we took into account effective $d-d$ hoppings ($t_\sigma$ and $t_\pi$ introduced correspondingly for the $a_{1g}$ and $e_g^\pi$ orbitals), trigonal crystal-field splitting and the spin-orbit coupling. The sketch of resulting level diagram is presented in Fig.~\ref{model-sketch}(a). In the left part of the figure we have purely MO picture with antibonding ($a$), nonbonding ($nb$), and bonding ($b$) orbitals. In the right part the situation with a very strong  spin-orbit coupling, which forms $j_{eff}^{1/2}$ and $j_{eff}^{3/2}$ levels is presented. Since the antibonding orbital of the $a_{1g}$ symmetry transforms to $j_{eff}^{1/2}$, while the one of $e_{g}^{\pi}$ symmetry into the $j_{eff}^{3/2}$ orbital the splitting between these antibonding states increases with $\lambda$ (for $\lambda \lesssim t_{\pi}$). This splitting calculated for parameters extracted from the Wannier function projection for Ba$_4$NbIr$_3$O$_{12}$ is shown in Fig.~\ref{model-sketch}(b). We see that the splitting grows with $\lambda$ exactly as in our GGA+SOC calculations, the results of which are presented in Figs.~\ref{Ir_SOC_bands},\ref{Ir_SOC_DOS}.

It is worth mentioning that the right part of Fig.~\ref{model-sketch}(a) represents an extreme situation of a very large spin-orbit coupling (comparing with $t_{\sigma}$, $t_{\pi}$, and  $\Delta_{CFS}^{a_{1g}-e_g^\pi}$) with atomic $j_{eff}^{1/2}$ and $j_{eff}^{3/2}$ levels. In real materials even in those of them based on $5d$ metals MOs are still formed.  The detailed analysis of the composition of the highest in energy $t_{2g}$ band shows that without SOC it is almost fully $a_{1g}$,  while introducing SOC leads to the mixing of ``pure'' orbitals and now it contains only 71$\%$ of the $a_{1g}$ orbital (it can be clearly seen from the comparison between total and $a_{1g}$ only density of states for Ir electrons both with and without including SOC, see Fig. \ref{part_dos_Ir}). This conclusion, that these are \textit{intersite effects} (i.e. hoppings) that can be detrimental to the famous $j_{eff}^{1/2}$ state of Ir$^{4+}$, was already formulated in Ref.~\cite{UFN,PNAS}, and the same conclusion was reached recently in the study of MO formation in lacunar spinel GaV$_4$S$_8$ in Ref.~{\cite{Haule}}.

The second evidence of the MO picture in Ba$_4$NbIr$_3$O$_{12}$ is a negligible value of magnetic moment per site. Even if we start from the magnetic spin-polarized situation (AFM, see above) our calculations converge to the nonmagnetic solution. It can be simply explained by the energy diagram of Fig.~\ref{levels}(a): 16 electrons per trimer fill all orbitals except the upper-lying antibonding $a_{1g}$ state, so that the ground state of Ba$_4$NbIr$_3$O$_{12}$ is just a nonmagnetic singlet. From this point of view it could be that the absence of magnetic ordering in this system is not due to the spin-liquid state anticipated in Ref.~\cite{Nguyen-2019-Ir}, but is just due to the formation of such nonmagnetic MO state.

However, experimental magnetic susceptibility in Ba$_4$NbIr$_3$O$_{12}$ follows the Curie-Weiss law with effective magnetic moment $\mu_{eff} = 0.8 \mu_B/f.u.$. Such a behaviour can be mimicked by the temperature-dependent Van Vleck paramagnetism. The singlet ground state in Ba$_4$NbIr$_3$O$_{12}$ shown in Fig.~\ref{levels}(a) may lie not far below the excited triplet state. In one-electron picture presented in Fig.~\ref{levels} this difference is an energy gap $\delta$ between anti-bonding $a_{1g}$ and $e_g^\pi$ states, but an account of many-body effects will strongly renormalize it giving $\delta^*$.  If $\delta^*$ is not so large $(\sim J_H)$ then the temperature fluctuations can throw one electron from the $e_g^{\pi}$ level to the highest $a_{1g}$ one and make up an excited triplet state with $S=1$. This energy can be rather small, see also discussion of the Rh case below.

Additionally, Van Vleck-like contributions can appear due to the spin-orbit coupling, since it generally does not commute with the term describing interaction with an external magnetic field $\hat H_B = \mu_B \vec B (g_s \hat {\vec S} + g_l \hat {\vec L})$. This results in mixing of different multiplets, which strongly affects temperature dependence of magnetic susceptibility. In fact recently these effects were shown to determine magnetic properties of dimerized materials with general formula A$_3$BTM$_2$O$_9$~\cite{Winter}.

\subsection{Ba$_4$NbRh$_3$O$_{12}$}
Finally, we discuss an ``intermediate'' situation represented by $4d$ transition metal Rh. In contrast to the case of  Ba$_4$NbIr$_3$O$_{12}$, according to our calculations the ground state of the Rh compound turns out to be ferromagnetic with the effective magnetic moment equal to 1.9 $\mu_B$ per trimer. It can be explained by the model of Fig.~\ref{levels} (c). If spitting $\delta$ between antibonding $a_{1g}$ and antibonding $e_g^\pi$ is not so large in comparison with $J_H$, it can lead to the Hund's energy gain exceeding  the energy splitting between antibonding $a_{1g}$ and $e_g^\pi$ levels  (note that the Hund's $J_H$ is larger for Rh than for Ir).  Then the ground state of Rh trimers could be the state shown in Fig.~\ref{levels}(c), i.e. it would correspond to the ``high-spin state'' of the Rh$_3$ molecule, with total spin $S=1$ and with the magnetic moment of 2$\mu_B$, in good correspondence with the results of our {\it ab initio} calculations. Thus the somewhat  stronger intra-atomic correlations  can make Rh trimer magnetic  - still retaining to a large extent the MO character, which one sees again by the strong bonding-antibonding splitting and in the shape of the (hole) density shown in Fig.~\ref{levels}(c) (cf. the Ir case in Fig.~\ref{levels}(b)). While spins within Rh$_3$ clusters are ferromagnetically ordered the superexchange coupling between trimers of $S=1$ is most likely antiferromagnetic. This would explain negative Curie-Weiss temperature seen experimentally~\cite{Nguyen-2019-Ir}. 

The COHP analysis for Rh trimer proves that it is an ``intermediate'' state between localized electron picture in $3d$ and MO model in $5d$. However, according to Fig.~\ref{COHP+U} the situation is closer to the Ir one as we still can see bonding and antibonding peaks, even though their magnitude is significantly lower. 

In effect we can conclude that the correlation effects arising when going from Ir to Rh, with trimers being crystallographically equivalent, lead to a magnetic solution for Ba$_4$NbRh$_3$O$_{12}$ with larger value of effective magnetic moment. Recently it has been also reported that another isostructural trimers made of Ru atoms that are also $4d$ but have another number of electrons per trimer, also have a magnetically-ordered ground state \cite{Nguyen-2018}. We have not done any calculations on this material. However, from the experimental findings its magnetic properties are rather similar to Rh - long-range magnetic ordering is observed below 4 K. We speculate that for both these $4d$ compounds MO picture is a good starting point but correlation effects should necessarily be taken into account.

\begin{figure}[t]
\includegraphics[width=0.48\textwidth]{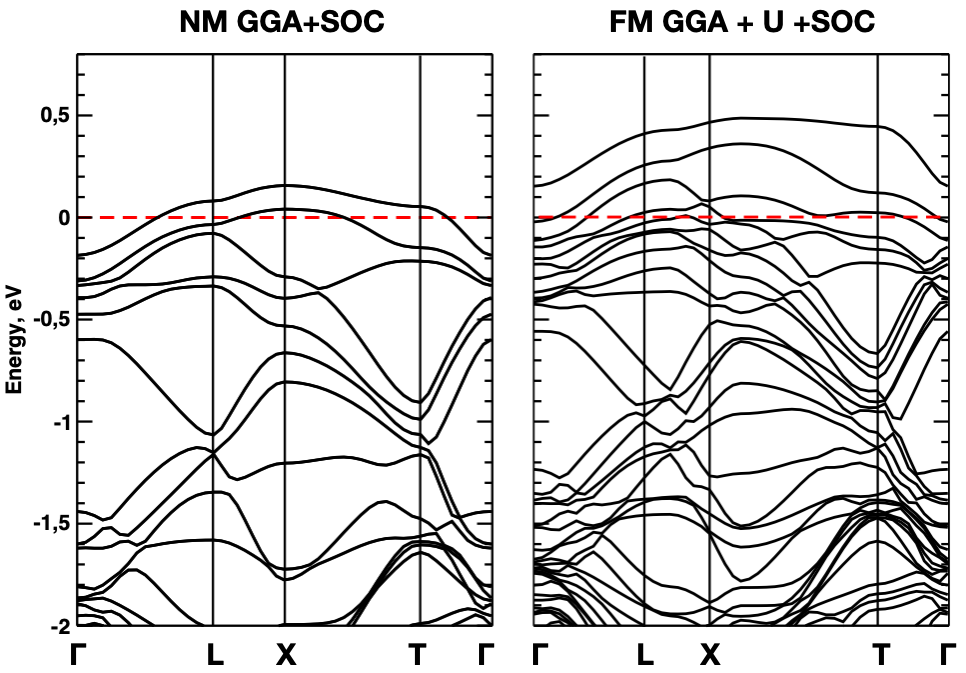}
\caption{\label{label} Calculated in nonmagnetic GGA+SOC (left) and magnetic GGA+U+SOC (right) band structures for Ba$_4$NbRh$_3$O$_{12}$. The Fermi energy is at zero. Note, in nonmagnetic GGA+SOC (left) all bands have additional spin-degeneracy.}
\label{Rh_SOC_bands}
\end{figure}

However, it should be mentioned that in contrast to the previously considered Ir-based material, we were not able to obtain insulating state in our calculations by introducing Hubbard's $U$ and Hund's $J_H$ in calculations. Account of the SOC shifts the highest band upwards (not shown), but this does not make our compound insulating as it is in the experiment. The resulting band structure for GGA+SOC and GGA+U+SOC calculations are shown in Fig.~\ref{Rh_SOC_bands}.

One of the possible explanations of metallic band structure is that the correlation effects in Ba$_4$NbRh$_3$O$_{12}$ should be considered in a more accurate way than in GGA+U(+SOC) method. While in Mn oxides one often even does not need to apply $U$ correction to obtain an insulating state (because of the very strong spin-splitting) and in Ba$_4$NbIr$_3$O$_{12}$, as we have shown, the SOC is large enough to open the gap, the situation in Ba$_4$NbRh$_3$O$_{12}$ can be more delicate one might need to apply more sophisticated methods such as, e.g., cluster version of dynamical mean-field theory (cluster DMFT)\cite{cluster-DMFT, CDMFT-1, CDMFT-2, CDMFT-3, CDMFT-4}. These dynamical correlations will renormalize the band dispersion and this may result in band gap opening. Moreover, they can be capable for formation of reduced (due to formation of MOs), but still localized magnetic moments and finally result in Curie behaviour. It could then be an example of a ``Curie metal''  (here rather small-gap Curie semiconductor) where strong correlations  should still be taken into account.


\section{Conclusions}
In summary, we considered three crystallographically equivalent compounds with structural $3d$, $4d$, and $5d$ TM trimers. By the direct calculations we confirmed the general tendency to have more correlated states in $3d$ systems and the picture that is closer to molecular description in $4d$ and especially $5d$ systems. Moreover, consideration of only Me-Me distance in not enough to prove formation of the molecular orbitals: while in Ba$_4$NbMn$_3$O$_{12}$ the Mn-Mn distance is smaller than in a pure metal, it does not form MOs. At the same time, in several compounds, e.g. lacunar spinels GaV$_4$S$_8$ or GeV$_4$S$_8$, MO picture works even though the V-V bond is significantly larger than in corresponding pure metals \cite{Haule}. Correct description of the electronic structure and magnetic properties for both Ir and Mn based compounds can be successfully done within correspondingly GGA+SOC and GGA+U approximations.


\section{Acknowledgements}
We are grateful to R.J. Cava for introducing us to trimer-based hexagonal perovskites and for various fruitful discussions. This research was supported by the Russian Foundation for Basic Researches (RFBR 20-32-70019) and the Russian Ministry of Science and High Education via program ``Quantum'' (No.\ AAAA-A18-118020190095-4) and contract 02.A03.21.0006 and the Deutche Forschungsgemeinschaft (DFG, German Reseach Foundation), project number 277146847-CRC 1238.


\end{document}